\documentclass[aps,pre,twocolumn,superscriptaddress,showpacs]{revtex4-1}
\usepackage{graphicx}
\usepackage{dcolumn}
\usepackage{bm}
\usepackage{epsfig}
\usepackage{color}

\begin{document}


\title{Incorporating a soft ordered phase into an amorphous
  configuration enhances its uniform plastic deformation under shear}


\author{Guo-Jie J. Gao}
\email{koh.kokketsu@shizuoka.ac.jp, gjjgao@gmail.com}
\affiliation{Department of Mathematical and Systems Engineering,
  Shizuoka University, Hamamatsu, Shizuoka 432-8561, Japan}
  
\author{Yun-Jiang Wang} \affiliation{State Key Laboratory of Nonlinear
  Mechanics, Institute of Mechanics, Chinese Academy of Sciences,
  Beijing 100190, China} \affiliation{School of Engineering Science,
  University of Chinese Academy of Sciences, Beijing 101408, China}

\author{Shigenobu Ogata} \affiliation{Department of Mechanical Science
  and Bioengineering, Osaka University, Toyonaka, Osaka 560-8531,
  Japan} \affiliation{Center for Elements Strategy Initiative for
  Structural Materials (ESISM), Kyoto University, Sakyo, Kyoto
  606-8501, Japan}

\date{\today}

\begin{abstract}
Amorphous materials of homogeneous structures usually suffer from
nonuniform deformation under shear, which can develop into shear
localization and eventually destructive shear band. One approach to
tackle this issue is to introduce an inhomogeneous structure
containing more than one phase, which can reduce the local nonuniform
shear deformation and hinder its percolation throughout the
system. Using thermostated molecular dynamics (MD) simulations, we
compare the deformation behavior between a homogeneous amorphous
mixture of bidisperse disc particles, interacting via an $n-6$
Lennard-Jones potential of tunable softness, with an inhomogeneous one
containing an evenly-distributed ordered phase. We change the
population ratio of large to small particles to create a homogeneous
or an inhomogeneous mixture, where the softness of a chosen phase can
be manually adjusted by specifying $n$ of the interparticle
potential. Results of applying extensive quasistatic shear on the
prepared mixtures reveal that the inhomogeneous amorphous mixture
containing a soft ordered phase overall deforms more uniformly than
the homogeneous one, which indicates that both the structure
inhomogeneity and the inter-phase softness variance play important
roles in enhancing the uniformity of the plastic deformation under
shear.
\end{abstract}


\maketitle

\section{Introduction}
\label{introduction}
Homogeneous amorphous materials such as bulk metallic glasses are
known for exhibiting superior mechanical properties than their
crystalline siblings. However, a significant disadvantage of amorphous
materials is their low ductility, due to nonuniform shear deformation,
which causes premature and unpredictable failure and greatly limits
their industrial applications \cite{greer06,falk16}. Recently,
introducing an ordered phase into an amorphous material to make the
engineered structure inhomogeneous have been shown to improve their
mechanical properties, for example, embedding an isolated and soft
crystal phase containing dendrites in a bulk metallic glass matrix to
enhance tensile ductility
\cite{johnson00,johnson01,johnson08_1,johnson08_2,liu12,zhang12}, and
distributing polycrystalline metallic alloys within amorphous shells
to raise mechanical strength and restrict shear localization
\cite{lu17}.

In this study, we propose a mesoscale model of mixing 2D bidisperse
disc particles to study the shear deformation behavior of an amorphous
configuration containing an ordered phase using molecular dynamics
(MD) simulations. In our model of 2D amorphous materials, we prepare a
homogeneous or an inhomogeneous amorphous/ordered configuration by
mixing $50-50$ or $90-10$ small and large $12-6$ Lennard-Jones (LJ)
disc particles, which allows the system to form an amorphous or a
partially amorphous structure under thermal equilibrium \cite{onuki06,
  onuki07}. We set the interparticle interaction to be Lennard-Jones
for the potential has long been used to study amorphous materials
including metallic glasses \cite{lund03,ogata06,ohern13}. We then
identify particles in the amorphous phase or the ordered phase based
on the values of their disorder parameter. We alter the strength of a
chosen phase by assigning the particles belonging to it proper
interparticle softness, using an $n-6$ Lennard-Jones (LJ) potential
\cite{shi11}. We choose $n=8$ to make particles in the ordered phase
softer than those in the amorphous phase. Finally, we apply
quasistatic shear on the prepared configurations and calculate the
uniformity of their deformation under extensive shear.

We investigated the difference in shear deformation between a
homogeneous and an inhomogeneous configurations and demonstrate that
an inhomogeneous configuration favors more uniform shear deformation
due to the spatial and strength heterogenity between the two
phases. Specifically, our model shows that introducing an ordered
phase into an amorphous phase helps the inhomogeneous system deform
more uniformly under shear than a homogeneous single-phased system
when the applied shear strain is small. For large shear strain, the
difference in softness between the two phases also plays an important
role to further enhance the uniform plastic deformation. Our MD
simulation results offer evidence supporting the improved ductility in
mesoscale reported in experiments of amorphous materials
\cite{johnson00,johnson01,johnson08_1,johnson08_2,liu12,zhang12,lu17}.

Below we elaborate on the 2D model of amorphous materials and the MD
simulation methods in section \ref{MD_method}, followed by
quantitative results of comparing the deformation behavior of
homogeneous and inhomogeneous configurations under quasistatic shear
in section \ref{uniformity of quasistatic shear deformation}. We
conclude our study in section \ref{discussions and conclusions}.

\section{Numerical simulation method}
\label{MD_method}
Our MD method includes two parts: 1) generating a homogeneous or
inhomogeneous initial amorphous configuration and 2) applying
quasistatic shear on it. To create an initial configuration, we first
pick up an equilibrium configuration in a liquid state where particles
interacting via the $12-6$ LJ potential and cool it down to a solid
temperature. Here temperature is defined as the total kinetic energy
divided by degrees of freedom of the system. Then we assign particles
different softness according to their disorder parameter and prepare
boundary particles. Finally, we relax the configuration sandwiched by
the boundary particles at the same solid temperature. To test the
deformation uniformity of the prepared initial configuration, we apply
quasistatic shear by moving the boundary particles stepwise followed
by thermostated relaxation and calculate the average deviation of
uniform deformation of the sheared configuration. For each system
setup, we use at least ten independent initial conditions to obtain
the averaged results. The details are given below.

\subsection{Preparation of an initial configuration}
\label{preparation of ICs}

\subsubsection{System geometry}
\label{System geometry}
Our system is a mixture of total $N=N_s+N_l$ circular particles
interacting via the finite-range, pairwise-additive LJ
potential. Specifically, it contains $N_s$ small particles of diameter
$d_s$ and $N_l$ large particles of diameter $d_l$, with the diameter
ratio $r=1.4$ to avoid artificial crystallization in 2D. We use
$N=1000$ throughout this study. The masses of the small particles
$m_s$ and the large particles $m_l$ are identical. The system occupies
a square simulation box of size $L$ on the $xy$-plane, where $x$ is
the horizontal axis and $y$ is the vertical one. For a given particle
number $N$ and particle diameters $d_s$ and $d_l$, the box size is
determined from the condition that the configuration has a fixed area
packing fraction $\phi=0.793$ or soft-core packing fraction of LJ
potential $\phi_s=1.0$. where $\phi$ and $\phi_s$ are defined as
\begin{equation} \label{packing fraction}
{\phi _s} = {{\pi \left[ {{N_s}{{\left( {\frac{{{2^{1/6}}{d_s}}}{2}}
            \right)}^2} + {N_l}{{\left( {\frac{{{2^{1/6}}{d_l}}}{2}}
            \right)}^2}} \right]} \mathord{\left/ {\vphantom {{\pi
          \left[ {{N_s}{{\left( {\frac{{{2^{1/6}}{d_s}}}{2}}
                  \right)}^2} + {N_l}{{\left(
                  {\frac{{{2^{1/6}}{d_l}}}{2}} \right)}^2}} \right]}
        {{L^2}}}} \right.  \kern-\nulldelimiterspace} {{L^2}}} =
{2^{1/3}}\phi.
\end{equation}
The value ${2^{1/6}}{d_s}$ or ${2^{1/6}}{d_l}$ is where the $12-6$ LJ
potential reaches its minimal, detailed below.

\subsubsection{Interparticle Lennard-Jones potential}
\label{interparticle potential}

We choose the finite-range, pairwise additive LJ potential to build
our amorphous model for it has been widely used to study amorphous
systems \cite{lund03,ogata06,ohern13}. To create a softness difference
between the amorphous and ordered phases in an inhomogeneous amorphous
configuration, we choose a tunable $n-6$ LJ potential \cite{shi11}
\begin{equation} \label{LJ_n_6_potential}
V_{LJ}^{n - 6}({r_{ij}}) = \left[ {4\epsilon \left( {\lambda
      {{(\frac{{{d_{ij}}}}{{{r_{ij}}}})}^n} - \alpha
      {{(\frac{{{d_{ij}}}}{{{r_{ij}}}})}^6}} \right) - {c_{ij}}}
  \right]\Theta (\frac{{{r_{cut}}}}{{{r_{ij}}}} - 1),
\end{equation}
where $\epsilon$ is the characteristic energy scale, $\lambda =
\frac{3}{2}[{{{2^{(n/6)}}} \mathord{\left/{\vphantom {{{2^{(n/6)}}}
          {(n - 6}}} \right.\kern-\nulldelimiterspace} {(n - 6}})]$,
$\alpha = n/[2(n - 6)]$, $r_{ij}$ is the separation between particles
$i$ and $j$, ${d_{ij}} = \frac{1}{2}({d_i} + {d_j})$ is their average
diameter, ${c_{ij}} = 4\epsilon [\lambda {({{{d_{ij}}}
      \mathord{\left/{\vphantom {{{d_{ij}}} {{r_{cut}}}}} \right.
        \kern-\nulldelimiterspace} {{r_{cut}}}})^n} - \alpha
  {({{{d_{ij}}} \mathord{\left/{\vphantom {{{d_{ij}}} {{r_{cut}}}}}
        \right.  \kern-\nulldelimiterspace} {{r_{cut}}}})^6}]$ is a
constant that guarantees $V_{LJ}^{n - 6}({r_{ij}}) \to 0$ as $r_{ij}
\to {r_{cut}}$, and $\Theta(x)$ is the Heaviside step function. We use
${r_{cut}} = 3.2{d_s}$, where $d_s$ is the diameter of small particles
and $n=8$ for the soft $V_{LJ}^{8 - 6}$ LJ potential or $n=12$ for the
well-known stiff $V_{LJ}^{12 - 6}$ LJ potential.

\subsubsection{Tuning inhomogeneity and softness}
\label{Tuning}
The whole process of preparing a homogeneous or inhomogeneous initial
configuration for our amorphous model can be summarized as follows. We
(a) create a randomly packed initial configuration of circular
particles without interparticle overlap; (b) equilibrate the system at
a liquid temperature $T_l$ to relax it; (c) pick up a relaxed liquid
configuration and equilibrate it again at a lower solid temperature
$T_s$; (d) attach boundary particles to the relaxed solid
configuration and reassign the interparticle interactions of all
particles based on the value of their disorder parameters; (e)
equilibrate the sandwiched solid configuration one last time at
$T=T_s$.

The MD simulations in this study use the diameter $d_s$ and mass $m_s$
of the small particles and the interparticle potential amplitude
$\epsilon$ as the reference length, mass, and energy scales. Mass
$m_l$ of the large particles is the same as $m_s$. As a result, the
unit of time $t$ is ${d_s}\sqrt {{m_s}/\epsilon}$. We control the
temperature $T$ of this system using the N{\'o}se-Hoover thermostat
with a thermostat moment of inertia $Q$ \cite{nose83, frenkel01}. Here
temperature $T$ is defined as the total kinetic energy
$\sum\nolimits_{i = 1}^N {{\raise0.7ex\hbox{${{m_i}v_i^2}$}
    \!\mathord{\left/ {\vphantom {{{m_i}v_i^2}
          2}}\right.\kern-\nulldelimiterspace}
    \!\lower0.7ex\hbox{$2$}}}$ divided by degrees of freedom of the
system and measured in units of $\epsilon/k_B$, where $m_i$ and $v_i$
are the mass and velocity of particle $i$, and $k_B$ is the
Boltzmann's constant. Degrees of freedom of the system is $2N-2$ or
$2N-1$ for periodic boundary conditions in both $x$ and $y$ directions
or only in the $x$ direction. To closely compare our results with
those in the literature \cite{onuki06, onuki07, onuki08, onuki10}, we
set $T=T_l=2.0$ and $T=T_s=0.2$ to study the system in a liquid state
or a solid state, respectively. The unit of $Q$ is $m_sd_s^2$. The
Newtonian equations of motion of a $N$ particle system excluding
boundary particles if any are integrated using the velocity Verlet
algorithm \cite{allen89}.

We perform a thermostated relaxation of the system excluding boundary
particles after generating a random non-overlapped initial
configuration and whenever making changes in the simulation
temperature, introducing boundary particles or reassigning properties
of particles for the pairwise $n-6$ LJ interparticle interactions. The
relaxation of a random initial configuration at $T=T_l$ also erases
any beginning memory from the system. We terminate the relaxation
process when the temperature fluctuation decays to within $\pm 5\%$ of
the required temperature.

It is known that by mixing bidisperse 2D particles with the population
ratio $c=N_l/N$ increasing from $0$ to $0.5$, we can create structure
from a single-crystal, a partially-amorphous (polycrystal) structure
to an amorphous structure \cite{onuki06, onuki07}. To create a
inhomogeneous amorphous configuration containing an ordered phase, we
choose a partially-amorphous structure using $c=0.1$ and $N=1000$.

After generating a random initial configuration (see the Appendix for
the implementing details), we equilibrate it at $T=T_l$. We then bring
the temperature down to $T_s$ within a time interval $\Delta t=10^5$
to equilibrium the system again with periodic boundary conditions in
both $x$ and $y$ directions and $Q=40.0$. Using the relaxed
configuration at $T=T_s$, we calculate the disorder parameter $D_j$ of
particle $j$ defined as \cite{nelson78, nelson02, onuki97, onuki98,
  onuki06, onuki07}
\begin{equation} \label{disorder parameter D_i}
{D_j} = 2\sum\limits_{k = 1}^{{N_b}} {[1 - \cos 6({\beta _j} - {\beta _k})]},
\end{equation}

where $\beta_j$ is a local crystalline angle introduced by
${e^{6i{\beta _j}}} = {{\sum\nolimits_{k = 1}^{{N_b}} {{e^{6i{\theta
            _{jk}}}}} } \mathord{\left/ {\vphantom {{\sum\nolimits_{k
            = 1}^{{N_b}} {{e^{6i{\theta _{jk}}}}} } {\left|
          {\sum\nolimits_{k = 1}^{{N_b}} {{e^{6i{\theta _{jk}}}}} }
          \right|}}} \right. \kern-\nulldelimiterspace} {\left|
    {\sum\nolimits_{k = 1}^{{N_b}} {{e^{6i{\theta _{jk}}}}} }
    \right|}}$, $i$ is the imaginary unit $\sqrt { - 1}$ and
$\theta_{jk}$ is the angle between the separation vector
${{\mathord{\buildrel{\lower3pt\hbox{$\scriptscriptstyle\rightharpoonup$}}
      \over r} }_j} -
{{\mathord{\buildrel{\lower3pt\hbox{$\scriptscriptstyle\rightharpoonup$}}
      \over r} }_k}$ of particle $j$ and its bonded neighboring
particle $k$ and the horizontal $x$ axis. Particle $k$ is considered
bonded to particle $j$ as long as $\left|
{{{\mathord{\buildrel{\lower3pt\hbox{$\scriptscriptstyle\rightharpoonup$}}
        \over r} }_j} -
  {{\mathord{\buildrel{\lower3pt\hbox{$\scriptscriptstyle\rightharpoonup$}}
        \over r} }_k}} \right| \le 1.5{d_{jk}}$ \cite{onuki07}. The
summation runs over all $N_b$ bonded neighboring particles $k$ of
particle $j$. The value of $D_j$ is zero if particle $j$ and its
bonded neighbors form an ordered perfect hexagonal structure and
increases if the structure becomes more disordered.

\begin{figure}
\includegraphics[width=0.40\textwidth]{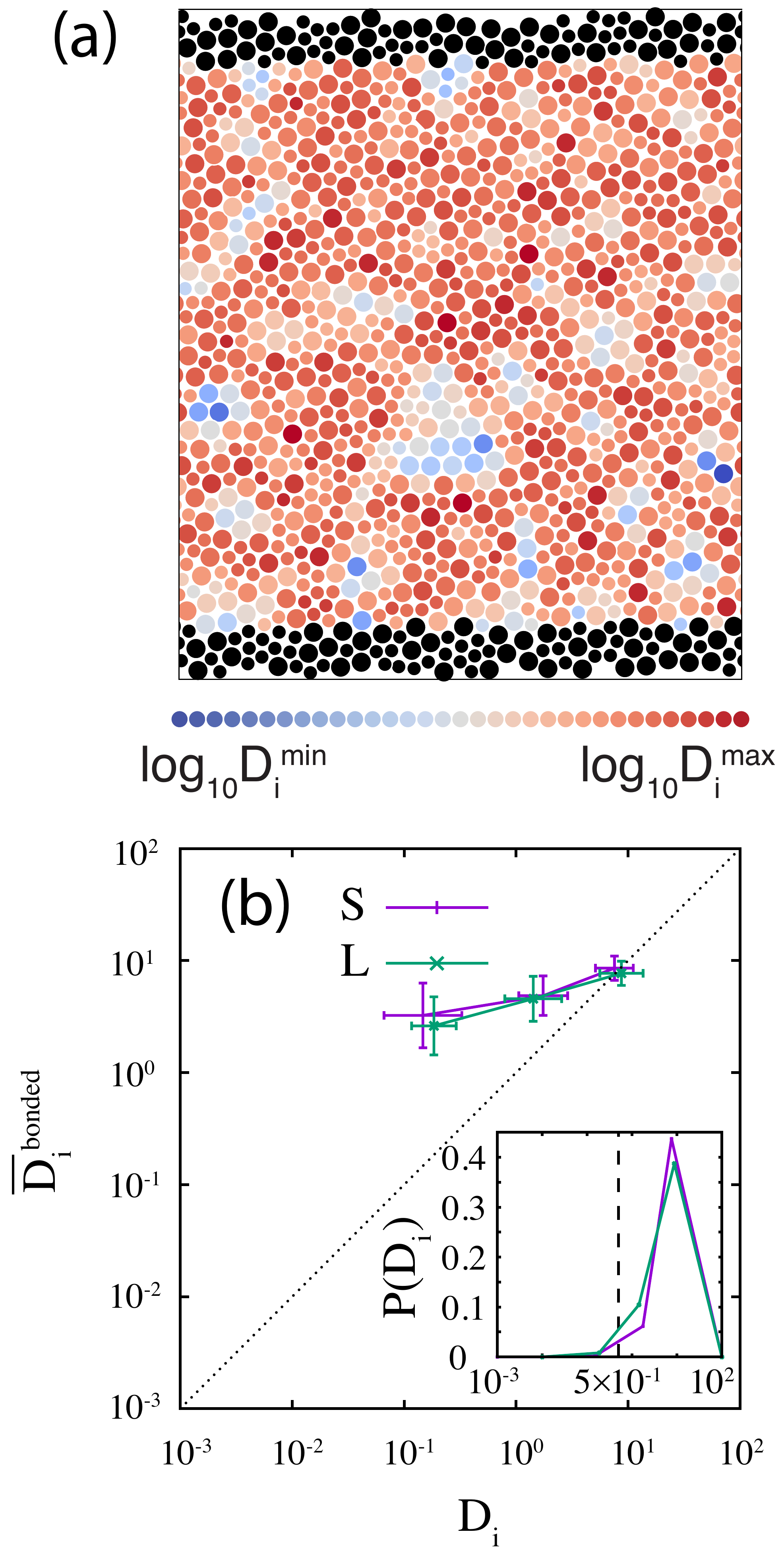}
\caption{\label{fig:AveNeighborD_c0.50} (color online) (a) A snapshot
  of an equilibrated configuration of $N=1000$ and $c=0.5$ at
  $T=T_s$. Particles are colored in shades from blue to red with
  increasing value of $log_{10}(D_i)$. The color scale is divided
  evenly between the minimal and maximal disorder parameters,
  $log_{10}(D_i^{min}) \approx -1.303$ and $log_{10}(D_i^{max})
  \approx 1.364$, in the system. Boundary particles are in black. (b)
  Averaged value of the disorder parameter of bonded neighbors of
  particle $i$ of disorder parameter value $D_i$. The inset shows the
  probability density distribution of $D_i$ with the same horizontal
  axis, and the total area below the two curves is one. The vertical
  dashed line shows a threshold $D_t=0.5$, separating ordered and
  disordered particles in this study. The data are obtained using ten
  relaxed configurations, where the results of small (S: purple) and
  large (L: green) particles are plotted separately.}
\end{figure}

Fig. \ref{fig:AveNeighborD_c0.50}(a) shows an equilibrated
fully-amorphous configuration of $c=0.5$ at $T=T_s$, where particles
are colored from blue to red with increasing
$log_{10}(D_i)$. Fig. \ref{fig:AveNeighborD_c0.50}(b) shows a weak
positive correlation between $D_i$ of particle $i$ and the average
$\bar D_i^{bonded}$ of the same quantity of its bonded neighbors,
obtained by averaging ten relaxed initial conditions. We can see that
particle $i$ can have ordered or disordered neighbors regardless of
its $D_i$. The figure shows two curves of $\bar D_i^{bonded}$, for
small or large particles, respectively. If we plot the probability
density function $P(D_i)$ of particle $i$ for small and large
particles separately, as shown in the inset of
Fig. \ref{fig:AveNeighborD_c0.50}(b), we can see quantitatively that
almost all particles are very disordered with $D_i>0.5$, showing the
configuration is indeed amorphous.

\begin{figure}
\includegraphics[width=0.40\textwidth]{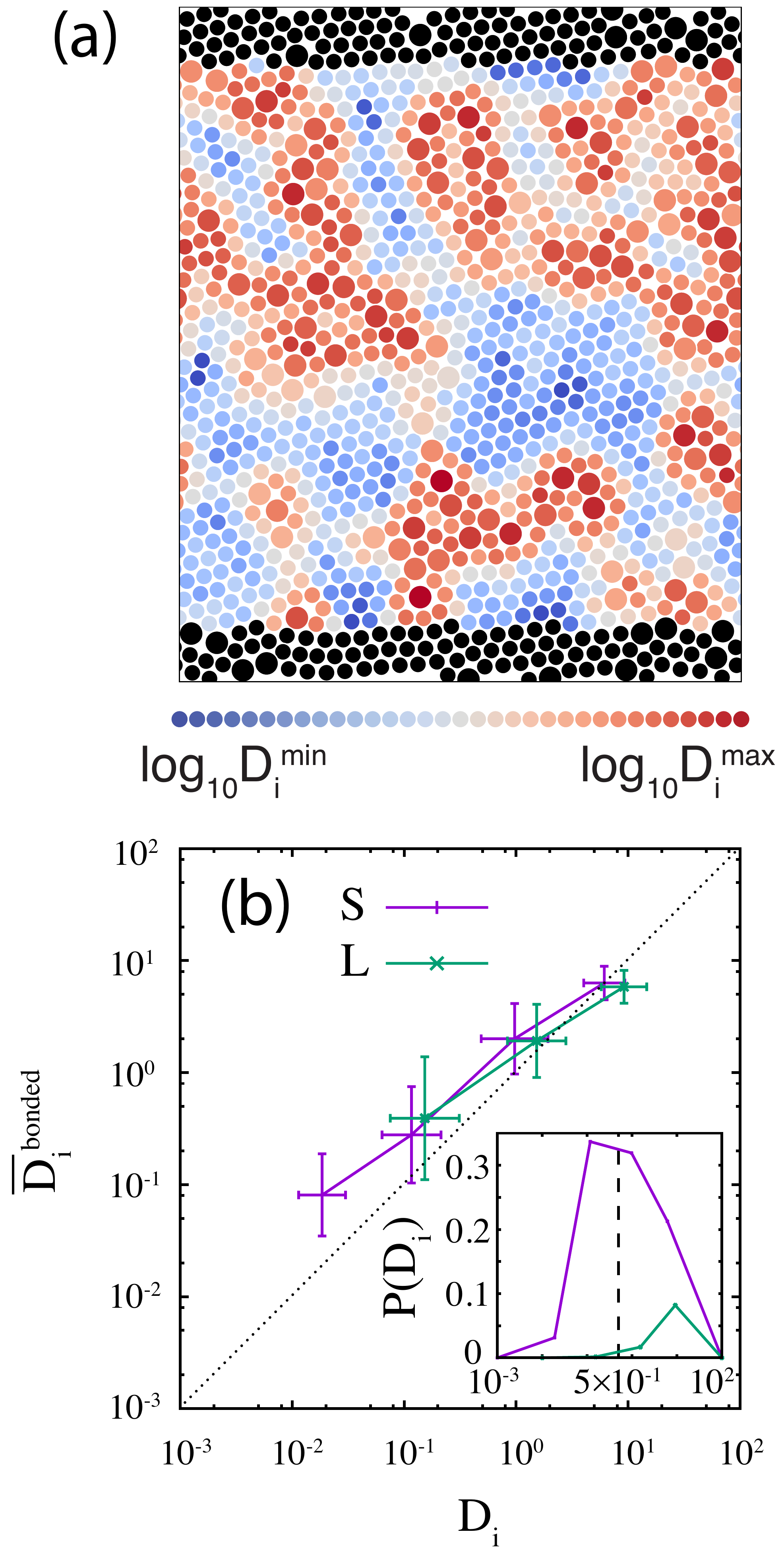}
\caption{\label{fig:AveNeighborD_c0.10} (color online) Same plots as
  in Fig. \ref{fig:AveNeighborD_c0.50}, except $c=0.10$. The color
  scale in (a) has a lower minimal and a similar maximal disorder
  parameters, $log_{10}(D_i^{min}) \approx -2.012$ and
  $log_{10}(D_i^{max}) \approx 1.411$, compared with those in
  Fig. \ref{fig:AveNeighborD_c0.50}(a).}
\end{figure}

Similarly, Fig. \ref{fig:AveNeighborD_c0.10} shows an equilibrated
partially-amorphous configuration of $c=0.1$ at $T=T_s$. We can
observe in Fig. \ref{fig:AveNeighborD_c0.10}(a) that a majority of
large particles are disordered and form the amorphous phase separating
ordered islands made of small particles. The most ordered small
particles form the cores of the ordered islands surrounded by less
ordered particles. The degree of disorder of particles increases
monotonically as a function of the distance measured from the ordered
cores. Contrary to Fig. \ref{fig:AveNeighborD_c0.50}(b),
Fig. \ref{fig:AveNeighborD_c0.10}(b) shows a strong positive
correlation between $D_i$ of particle $i$ and its $\bar D_i^{bonded}$,
also using ten relaxed initial conditions. We can clearly see that
very ordered particles tend to have ordered neighbors, but highly
disordered particles tend to have equally disordered neighbors, which
gives the two curves positive slopes close to unity. Moreover, in the
inset of Fig. \ref{fig:AveNeighborD_c0.10}(b), we can see
quantitatively that almost all large particles are very disordered
with $D_i>0.5$, while only about half small particles are in a similar
disordered status.

The insets of Fig. \ref{fig:AveNeighborD_c0.50} and
Fig. \ref{fig:AveNeighborD_c0.10} disclose the basic principle of how
to build an homogeneous or inhomogeneous amorphous structure out of a
mixture of small and large particles in this study: in a
fully-amorphous configuration, all most all particles have their
$D_i>D_t$, where $D_t=0.5$ is a threshold disorder value used in this
study. On the other hand, in a partially-amorphous configuration, we
can use $D_i<D_t$ to identify ordered particles, which form an ordered
phase, and the rest particles form an amorphous phase. Of the ten
initial configurations used in this study, the average number of
ordered particles is $0.4406N$. The amorphous phases in homogeneous
and inhomogeneous configurations are similar in terms of their
disorder parameter distributions.

\begin{figure}
\includegraphics[width=0.40\textwidth]{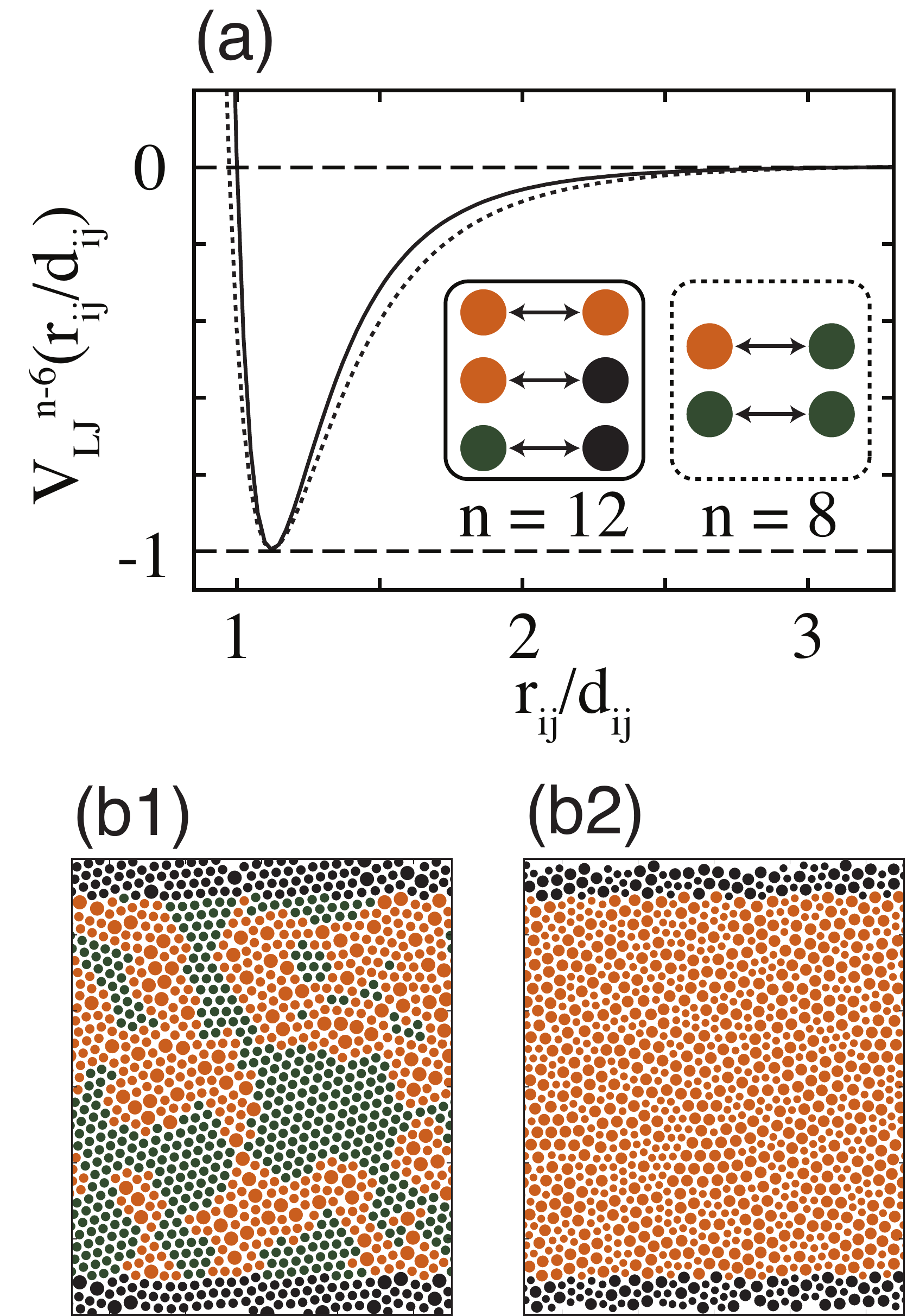}
\caption{\label{fig:scheme} (color online) The $n-6$ LJ interparticle
  potential used in the 2D model of amorphous materials, shown in
  (a). The interaction between a soft ordered (green) particle and
  another soft ordered or stiff disordered (orange) particle is the
  soft $V_{LJ}^{8-6}$ LJ potential. The interaction between two stiff
  disordered particles or with a boundary particle (black) is the
  stiff $V_{LJ}^{12-6}$ LJ potential. Exemplary snapshots of an
  inhomogeneous and a homogeneous configurations are shown in (b1) and
  (b2), respectively.}
\end{figure}

For an inhomogeneous amorphous configuration, we assign the soft
interparticle $V_{LJ}^{8-6}$ potential to the interaction between two
ordered particles of ${D_i} \le {D_t}$. We assign the same soft
$V_{LJ}^{8-6}$ potential to the interaction between an ordered
particle of ${D_i} \le {D_t}$ and a disordered particle of ${D_i} >
{D_t}$, happening at the interface between the ordered and amorphous
phases. Moreover, the stiff interparticle $V_{LJ}^{12-6}$ potential
governs the interaction between two disordered particles of ${D_i} >
{D_t}$. Finally, the interaction between an ordered or disordered
particle and a boundary particle is controlled by the same stiff
$V_{LJ}^{12-6}$ potential. The boundary particles are image particles
created using the periodic boundary condition in the vertical $y$
axis. We keep enough image particles so that the top and bottom
boundaries have thickness equals $r_{cut}=3.2d_s$, and the total
number of top or bottom boundary particles is about $3.2{d_s}\sqrt
N$. Boundary particles are not thermostated, and we relax the prepared
system sandwiched by boundary particles at $T=T_s$ again with periodic
boundary condition in the horizontal $x$ direction and $Q=0.01$ before
using it for the quasistatic shear tests. We observe only slight local
position variation but no large-scale rearrangement of particles in
the amorphous or ordered phase of the relaxed configuration. Two
equilibrated snapshots of exemplary inhomogeneous and homogeneous
amorphous configurations and the rules for interparticle interactions
are shown in Fig. \ref{fig:scheme}, where particles in soft-ordered
and stiff-amorphous phases are colored in green and orange,
respectively.

\subsection{Quasistatic shear deformation}
\label{quasistatic shear}

\begin{figure}
\includegraphics[width=0.40\textwidth]{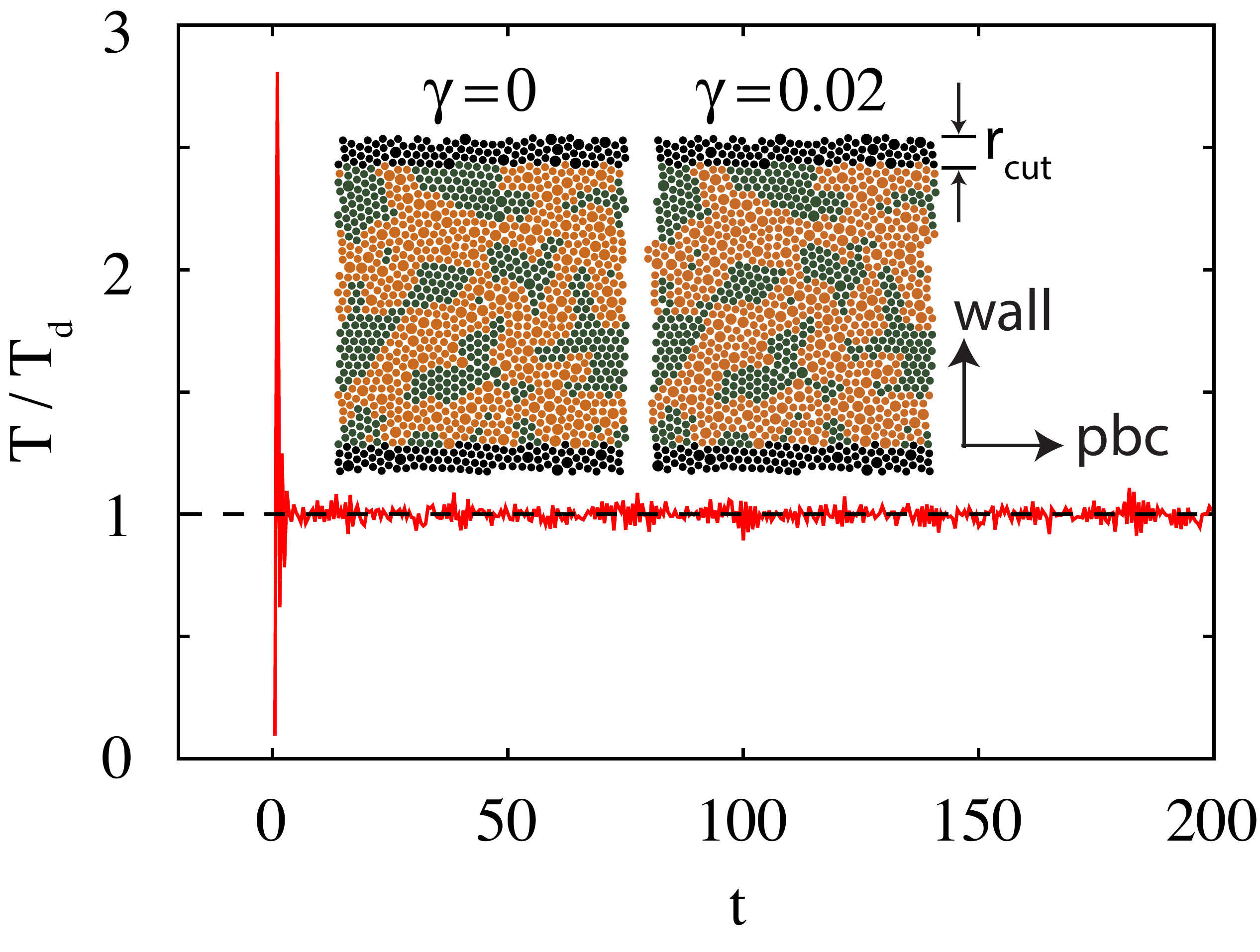}
\caption{\label{fig:quasistatic_shear} (color online) A sudden
  increase in temperature $T$ and its fluctuation, as a function of
  time $t$, corresponding to an increment of shear strain $\gamma$ of
  the system by $0.02$ in the quasistatic shear deformation. The
  quasistatic shear is done by evenly shifting the top and bottom
  boundary particles, within a thickness of $r_{cut}$, to the opposite
  directions, respectively. The perturbed temperature decays quickly
  to the desired value $T_d=T_s$ within the observation time
  interval.}
\end{figure}

We apply quasistatic shear strain on the prepared homogeneous and
inhomogeneous amorphous configurations to test their plastic
deformation behavior. To do this, at each step of the quasistatic
shear, we shift the $x$ position of each top and bottom boundary
particle by a small amount of $0.01L$ and $-0.01L$, respectively. The
system heats up due to the perturbation from the moved boundary
particles. Using a dimensionless MD time step $dt=0.001$ and the
periodic boundary condition in the horizontal $x$ direction, we relax
the system at temperature $T_s$ with the N{\'o}se-Hoover thermostat of
$Q=0.01$ integrated by the velocity Verlet algorithm until the
standard deviation of temperature $T$, periodically calculated within
a time interval of $2,500$ MD steps, decays to smaller than $0.03$,
which corresponds to a temperature fluctuation within $\pm 5\%$ of the
assigned temperature. Boundary particles are not thermostated and
their positions stay fixed during the relaxation
process. Fig. \ref{fig:quasistatic_shear} demonstrates the decay of
temperature fluctuation while applying this process on an exemplary
inhomogeneous configuration. We repeat the two steps of shifting
boundary particles and relaxation of the system at $T_s$ until a
prescribed value of shear strain $\gamma$ is reached.

\section{Analysis of the uniformity of deformation under quasistatic shear}
\label{uniformity of quasistatic shear deformation}

\begin{figure}
\includegraphics[width=0.40\textwidth]{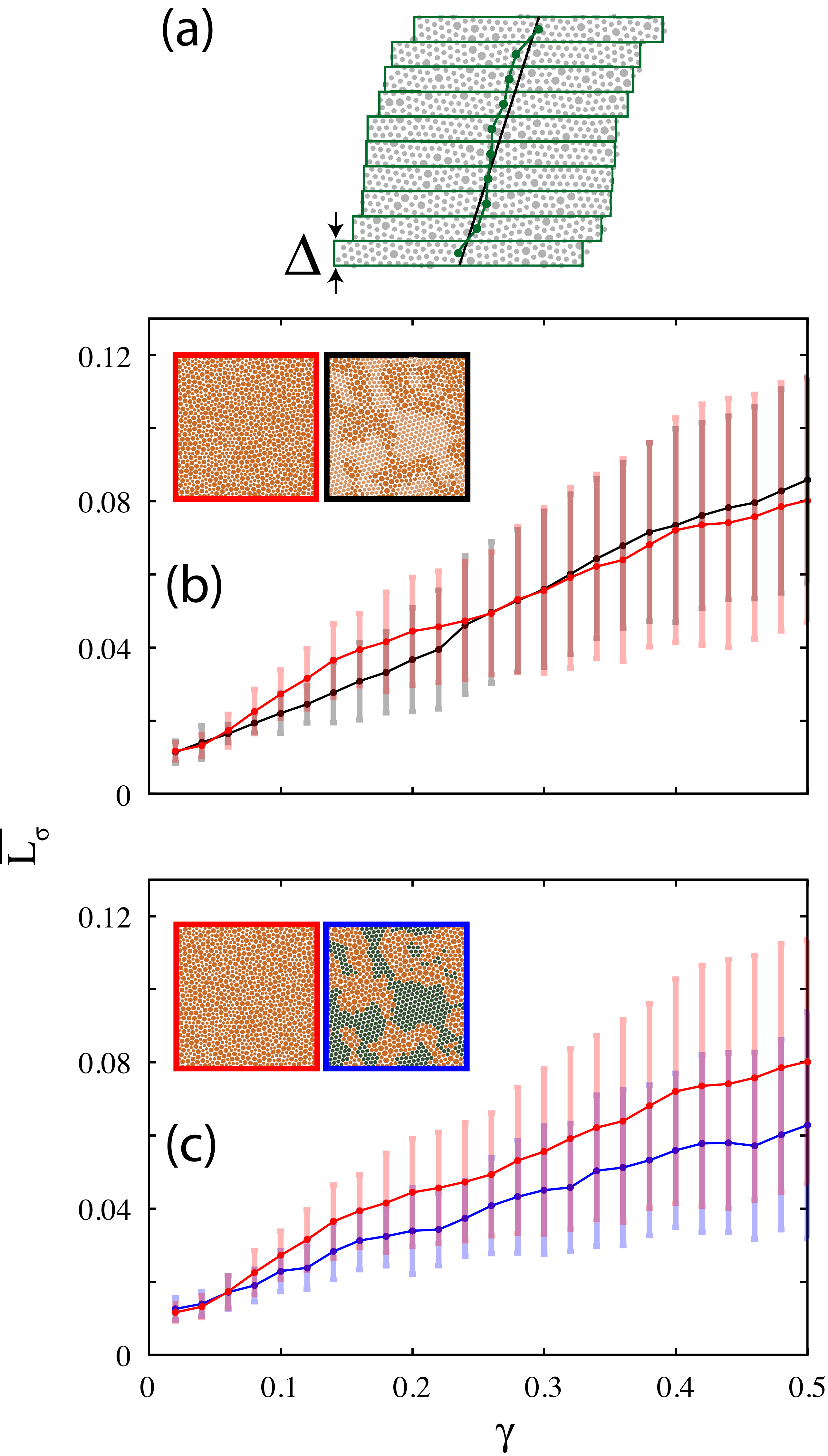}
\caption{\label{fig:uniformity_summary_T2d-1_10trials} (color online)
  (a) The definition of the deviation of uniform deformation
  $L_{\sigma}$ which measures the difference between the averaged $x$
  positions of $m=10$ sliced sections (connected green line), each
  with an identical thickness $\Delta$, of a given configuration and
  the $x$ positions of its calculated counterpart from the idealized
  uniformly-deformed profile (black line). (b) Averaged $L_{\sigma}$
  of homogeneous configurations (red) and inhomogeneous configurations
  without softness inhomogeneity (black). In the inset, particles that
  are more ordered of an inhomogeneous configuration are
  semi-transparent. (c) Averaged $L_{\sigma}$ of the same homogeneous
  configurations (red) and the same inhomogeneous configurations
  except with softness inhomogeneity (blue). In the insets, the stiff
  and soft phases are colored in orange and green, respectively. For
  each case, the results are obtained using ten initial
  configurations.}
\end{figure}

As mentioned in the Introduction, we expect that an inhomogeneous
amorphous configuration allows the system to shear more uniformly than
a homogeneous one for the latter lack of the inhomogeneity of two
phases with different softness to deter the emergence of shear
localization. To measure the uniformity of shear deformation
quantitatively, we calculate the deviation of uniform shear
deformation $L_\sigma$ that measures how far a sheared configuration
at a given strain $\gamma$ is away from its completely uniformly
deformed counterpart, defined as
\begin{equation} \label{L_sigma}
{L_\sigma }(\gamma ) = \sqrt {\frac{{\sum\nolimits_{i = 1}^m
      {{{[{{\bar x}_i}(\gamma )\left| {_{y \in [{y_i}(\gamma ) -
                    {\Delta \mathord{\left/ {\vphantom {\Delta 2}}
                        \right. \kern-\nulldelimiterspace}
                      2},{y_i}(\gamma ) + {\Delta \mathord{\left/
                        {\vphantom {\Delta 2}}
                        \right. \kern-\nulldelimiterspace} 2}]}}
              \right. - {x_i}(\gamma )]}^2}} }}{m}},
\end{equation}
where $x_i$ and $y_i$ are calculated positions of the idealized
uniform deformation profile, evenly divided into $m$ equal-sized
stripes, $\Delta$ is the width of each stripe, and ${{{\bar x}_i}}$ is
the averaged $x$ positions of particles whose $y$ positions located
within ${y_i} - \frac{\Delta }{2}$ and ${y_i} + \frac{\Delta }{2}$ of
stripe $i$. To make sure that the choice of $m$ has no influence on
our conclusion, we have tried $m = 5, 10, 20, 30, 40$ and $50$ and
found that the values of $L_\sigma$ stay unchanged within negligible
fluctuations. We therefore use $m=10$ throughout our analysis and ten
trials with different initial conditions to calculate an averaged
${{\bar L}_\sigma }$ as a function of $\gamma$, as shown in
Fig. \ref{fig:uniformity_summary_T2d-1_10trials}(a).

Our amorphous model with tunable inhomogeneity allows us to compare
the effects of introducing an inhomogeneous configuration and creating
a softness difference between the two phases in the inhomogeneous
configuration on the uniformity of shear deformation independently. We
proceed with our investigation by two stages. First, to test the
effect of introducing the inhomogeneous configuration, we compare the
shear deformation behavior between an amorphous homogeneous
configuration and a partially-amorphous inhomogeneous configuration,
where the interparticle interactions are $V_{LJ}^{12 - 6}$ LJ
potential in both cases. The results of the comparison is shown in
Fig. \ref{fig:uniformity_summary_T2d-1_10trials}(b). We can see
clearly that introducing a configurational inhomogeneity effectively
reduces nonuniform deformation when strain $\gamma$ is smaller than
about $0.2$. When sheared further, the inhomogeneous configuration
gradually loses to the homogeneous one.

Second, we incorporate the softness difference between the two phases
of the inhomogeneous configuration, where now ordered particles
interacting with another ordered or disordered particles via the soft
$V_{LJ}^{8 - 6}$ LJ potential, and compare it with the homogeneous one
as in the first stage. The results are shown in
Fig. \ref{fig:uniformity_summary_T2d-1_10trials}(c). Strikingly, the
tuned inhomogeneous configuration deforms even more uniformly (smaller
averaged $L_\sigma$) than the homogeneous one until the shear strain
$\gamma$ reaches about $0.5$.

\begin{figure}
\includegraphics[width=0.40\textwidth]{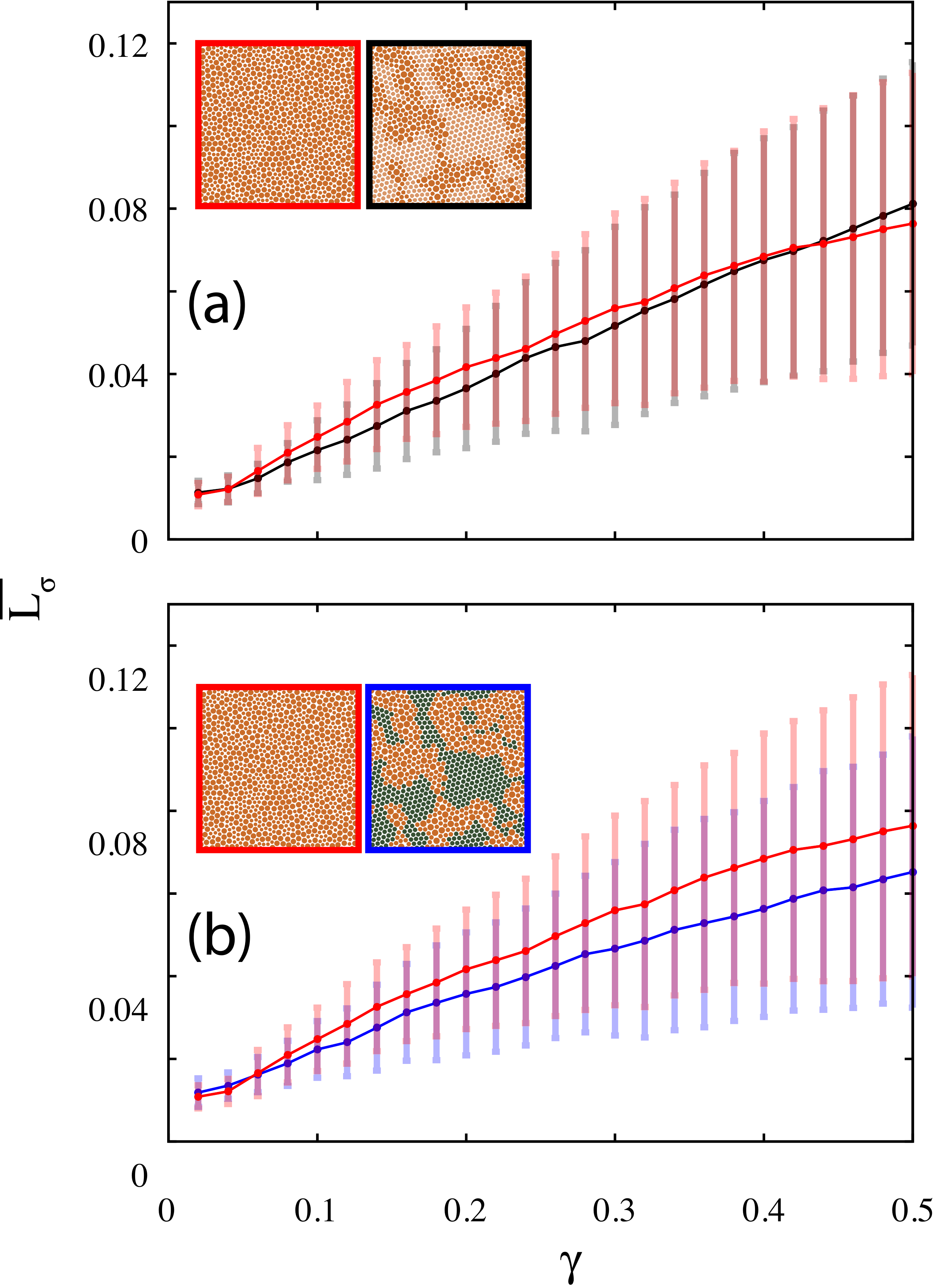}
\caption{\label{fig:uniformity_summary_T2d-1_50trials} (color online)
  Averaged $L_{\sigma}$ as in
  Fig. \ref{fig:uniformity_summary_T2d-1_10trials} (b) and (c), except
  the results are obtained using fifty initial configurations for both
  cases.}
\end{figure}

Due to the large error bars in
Fig. \ref{fig:uniformity_summary_T2d-1_10trials}(b) and (c), we
further verify the results using five times more initial
configurations to verify the reliability of the observed trends. We
confirm the trends stay the same and therefore saturate the data, as
shown in Fig. \ref{fig:uniformity_summary_T2d-1_50trials}.

\begin{figure}
\includegraphics[width=0.40\textwidth]{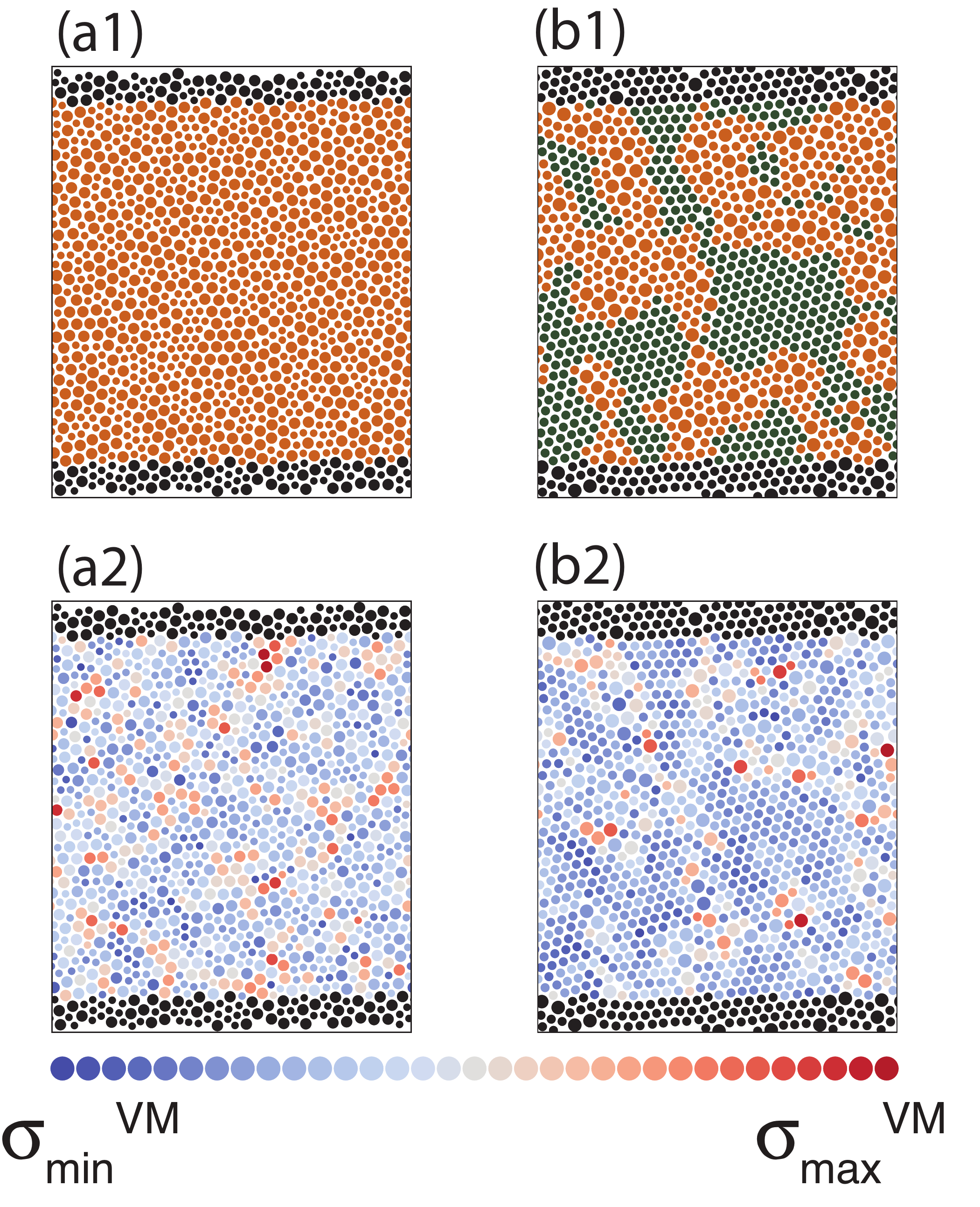}
\caption{\label{fig:type_vonMisesP_compare} (color online) Snapshots
  of an equilibrated (a1) homogeneous and (b1) inhomogeneous
  configurations of $N=1000$ at $T=T_s$ and $\gamma=0.02$, colored by
  particle types (soft ordered: green; stiff disordered: orange)
  defined in the initial conditions. $c=0.1$ for the inhomogeneous
  configuration. Particles are colored in shades from blue to red with
  increasing value of the von Mises stress $\sigma^{VM}$ in (a2) and
  (b2) correspondingly. The color scale is relative to the minimal and
  maximal von Mises stresses $\sigma_{min}^{VM}$ and
  $\sigma_{max}^{VM}$ in each system.}
\end{figure}

\begin{figure}
\includegraphics[width=0.40\textwidth]{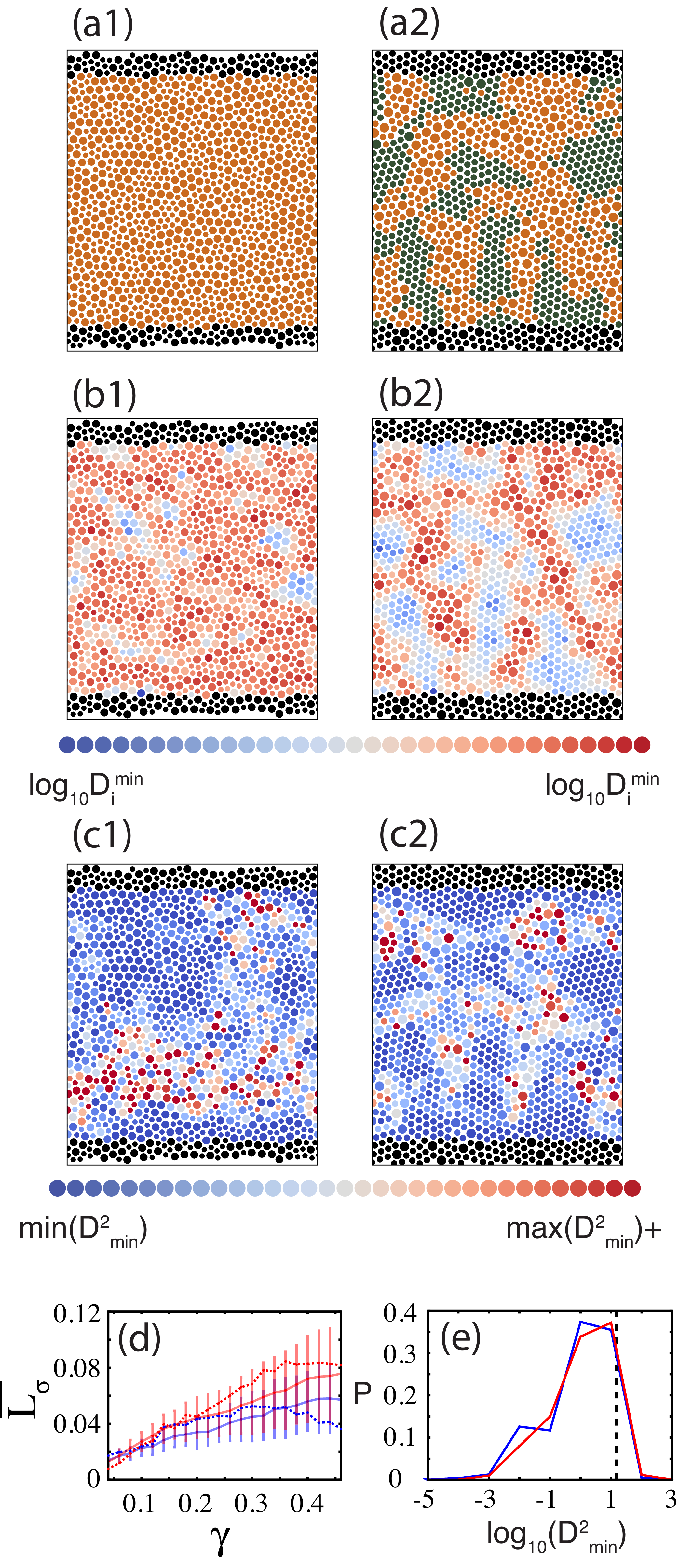}
\caption{\label{fig:D_min2_comparison} (color online) Snapshots of
  sheared homogeneous (a1) and inhomogeneous (a2) configurations at
  $\gamma=0.46$, colored by particle types (soft ordered: green; stiff
  disordered: orange). Particles of the homogeneous and inhomogeneous
  configurations are colored by the disorder parameter $D_i$ in (b1)
  with $log_{10}(D_i^{min}) \approx -1.544$ and $log_{10}(D_i^{max})
  \approx 1.367$, and in (b2) with $log_{10}(D_i^{min}) \approx
  -2.179$ and $log_{10}(D_i^{max}) \approx 1.369$, respectively. (c1)
  and (c2) show particles colored by $D_{min}^2$ with a sampling
  radius of $1.49d_s$ and a reference configuration at $\gamma = 0.04$
  after eliminating homogeneous cell deformation. The color scale is
  evenly divided between $min(D_{min}^2)=0$ and $max(D_{min}^2)=15$,
  and particles having $D_{min}^2 \geq max(D_{min}^2)$ (homogeneous:
  $6.9\%$; inhomogeneous: $4.2\%$) are colored by the darkest
  red. ${{\bar L}_\sigma }$ of the homogeneous (dotted red) and
  inhomogeneous (dotted blue) configurations within
  $\gamma=[0.04,0.46]$ is shown in (d), having the averaged ${{\bar
      L}_\sigma }$ in
  Fig. \ref{fig:uniformity_summary_T2d-1_10trials}(c) for
  reference. The corresponding probability density distribution
  $P(log_{10}(D_{min}^2))$, with $max(D_{min}^2)$ indicated by the
  dashed line, is shown in (e).}
\end{figure}

To get a better insight into the deformation mechanism under shear of
the homogeneous and inhomogeneous configurations, we calculate their
per-particle von Mises stress, which is the deviatoric part of the
stress tensor, defined as $\sigma _i^{VM} = \sqrt {{{(\sigma
      _i^{xx})}^2} + {{(\sigma _i^{yy})}^2} - \sigma _i^{xx}\sigma
  _i^{yy} + 3{{(\sigma _i^{xy})}^2}}$ in 2D, where $\sigma _i^{ab} = -
[m_iv_i^av_i^b + \sum\nolimits_{j = 1}^{{N_p}}
  {\frac{1}{2}r_{ij}^af_{ij}^b} ]$ for particle $i$ influenced by
$N_p$ neighbors within ${r_{cut}}$ via the $n-6$ Lennard-Jones
potential, $f_{ij}$ is the force acting on particle $i$ from its
neighbor $j$, and $a,b \in [x,y]$ \cite{thompson09}. The results are
shown in Fig. \ref{fig:type_vonMisesP_compare}. We observe that in a
inhomogeneous configuration, particles subject to high $\sigma^{VM}$
shear stress mostly distribute within the amorphous phase, and the
isolated soft-ordered phase disperses them so that they cannot
coordinate to percolate through the whole system easily. Furthermore,
we also calculate the local deviation from affine deformation,
$D^2_{min}$, which identifies local irreversible particle shuffling in
unit of $d_s^2$ \cite{falk98}. We show a comparison of $D^2_{min}$
between exemplary homogeneous and inhomogeneous configurations during
the quasistatic shear strain interval $\gamma = [0.04, 0.46]$ in
Fig. \ref{fig:D_min2_comparison}. Both systems have similar
probability density distribution $P(log_{10}(D_{min}^2))$ and a
similar amount of particles are subject to irreversible shear
deformations equal or greater than the same $max(D^2_{min})$ at
$\gamma=0.46$. We can see clearly that particles with large
$D^2_{min}$ concentrate on one side in the homogeneous system,
responsible for the higher nonuniform deformation. On the other hand,
similar particles are mostly distributed evenly within the amorphous
phase, and the intermediate ordered phase hinders their percolation,
which enhances the uniform deformation. Our calculations of $\sigma
_i^{VM}$ and $D^2_{min}$ offer clear evidence that an inhomogeneous
amorphous configuration can effectively deter nonuniform shear
deformation than a homogeneous one.

\section{Conclusions}
\label{discussions and conclusions}

In this study, we propose a 2D mixture of bidisperse particles to
study the deformation behavior of homogeneous and inhomogeneous
amorphous materials. A configuration modeling homogeneous amorphous
materials is made of $50-50$ small and large particles. On the other
hand, Our mesoscale model of inhomogeneous amorphous materials
distributes large circular particles in a sea of small particles with
a $1:9$ population ratio. In the inhomogeneous model, about half
($\approx 44\%$) small particles form an isolated and ordered
phase. On the other hand, most large particles and the remaining half
small particles together form an amorphous phase filling the space in
between the ordered phase. To give the ordered or the amorphous phase
proper mechanical softness as proposed in experiments on amorphous
materials to improve their ductility, we introduce a tunable $n-6$ LJ
potential. A particle in the soft-ordered phase interacts with another
particle via the $8-6$ LJ potential. Two disordered particles interact
with each other via the stiff $12-6$ LJ potential. We apply
quasistatic shear on the homogeneous and inhomogeneous configurations
by repeatedly moving the boundary particles with a stepwise shear
strain of $0.02$, followed by thermostated relaxation.

Our simulation results show that when the applied shear strain
$\gamma$ is small, the configurational inhomogeneity between the two
phases in an inhomogeneous structure alone works well to improve
uniform deformation. However, with increasing the shear deformation
further, the difference in softness between the two phases plays an
essential role to enhance the uniform shear behavior. In general, the
inhomogeneous configurations deform more uniformly than the
homogeneous ones. Our simplified model offers clear numerical evidence
supporting the experimental results and could open a new quantitative
approach to systematically improving the ductility of inhomogeneous
amorphous materials including an ordered phase. For the future work,
we will examine a hard ordered phase embedded in a soft disordered
matrix and explore the optimal ordered/disordered area ratio that
gives the best deformation uniformity, which will give a more complete
picture of this study.

\section{Appendix: Generating a randomly packed configuration of particles}
\label{Appendix}

When conducting MD simulation in a liquid state, we start with an
initial configuration at $\phi=0.793$ without overlap between
particles. Practically, it is very difficult to generate such random
initial configuration using a completely random process which places
particles one by one. To avoid this issue, we first generate a
mechanically stable (MS) packing of frictionless particles interacting
via the finite-range pairwise-additive repulsive spring potential at
area packing fraction $\phi=0.84$ ($\phi_s=1.06$), close to
random-close packing density in 2D.

The MS packing of particles is generated using the procedure detailed
in \cite{gao06}. We start with a sparse initial configuration, and the
procedure increases the sizes of the particles followed by energy
minimization to remove overlap between particles. Periodic boundary
conditions are implemented in both $x$ and $y$
directions. Occasionally, particles have to be shrunk if the energy
minimization procedure fails to remove interparticle overlap. We
repeat the two steps of particle size perturbation and energy
minimization until all particles are force-balanced with their
neighbors, and any attempt to increase particle size will result in an
increase of the total energy of the system.

We then decrease $\phi$ from about $0.84$ ($\phi_s=1.06$) to $0.793$
($\phi_s=1.0$) manually and run the thermostated MD simulation at
$T=T_l$. Since the system is in a liquid state, it will soon forget it
was from an MS state as long as it is fully relaxed.

\section{acknowledgments}
GJG acknowledges financial support from Shizuoka University startup
funding. YJW acknowledges financial support from National Natural
Science Foundation of China (NSFC No.11672299). We also thank Corey
S. O'Hern for insightful comments.

\bibliography{paper2}

\providecommand{\noopsort}[1]{}\providecommand{\singleletter}[1]{#1}%
\begin{thebibliography}{27}%
\makeatletter
\providecommand \@ifxundefined [1]{%
 \@ifx{#1\undefined}
}%
\providecommand \@ifnum [1]{%
 \ifnum #1\expandafter \@firstoftwo
 \else \expandafter \@secondoftwo
 \fi
}%
\providecommand \@ifx [1]{%
 \ifx #1\expandafter \@firstoftwo
 \else \expandafter \@secondoftwo
 \fi
}%
\providecommand \natexlab [1]{#1}%
\providecommand \enquote  [1]{``#1''}%
\providecommand \bibnamefont  [1]{#1}%
\providecommand \bibfnamefont [1]{#1}%
\providecommand \citenamefont [1]{#1}%
\providecommand \href@noop [0]{\@secondoftwo}%
\providecommand \href [0]{\begingroup \@sanitize@url \@href}%
\providecommand \@href[1]{\@@startlink{#1}\@@href}%
\providecommand \@@href[1]{\endgroup#1\@@endlink}%
\providecommand \@sanitize@url [0]{\catcode `\\12\catcode `\$12\catcode
  `\&12\catcode `\#12\catcode `\^12\catcode `\_12\catcode `\%12\relax}%
\providecommand \@@startlink[1]{}%
\providecommand \@@endlink[0]{}%
\providecommand \url  [0]{\begingroup\@sanitize@url \@url }%
\providecommand \@url [1]{\endgroup\@href {#1}{\urlprefix }}%
\providecommand \urlprefix  [0]{URL }%
\providecommand \Eprint [0]{\href }%
\providecommand \doibase [0]{http://dx.doi.org/}%
\providecommand \selectlanguage [0]{\@gobble}%
\providecommand \bibinfo  [0]{\@secondoftwo}%
\providecommand \bibfield  [0]{\@secondoftwo}%
\providecommand \translation [1]{[#1]}%
\providecommand \BibitemOpen [0]{}%
\providecommand \bibitemStop [0]{}%
\providecommand \bibitemNoStop [0]{.\EOS\space}%
\providecommand \EOS [0]{\spacefactor3000\relax}%
\providecommand \BibitemShut  [1]{\csname bibitem#1\endcsname}%
\let\auto@bib@innerbib\@empty
\bibitem [{\citenamefont {Ashby}\ and\ \citenamefont {Greer}(2006)}]{greer06}%
  \BibitemOpen
  \bibfield  {author} {\bibinfo {author} {\bibfnamefont {M.}~\bibnamefont
  {Ashby}}\ and\ \bibinfo {author} {\bibfnamefont {A.}~\bibnamefont {Greer}},\
  }\href@noop {} {\bibfield  {journal} {\bibinfo  {journal} {Scripta Mater.}\
  }\textbf {\bibinfo {volume} {54}},\ \bibinfo {pages} {321} (\bibinfo {year}
  {2006})}\BibitemShut {NoStop}%
\bibitem [{\citenamefont {Hufnagel}\ \emph {et~al.}(2016)\citenamefont
  {Hufnagel}, \citenamefont {Schuh},\ and\ \citenamefont {Falk}}]{falk16}%
  \BibitemOpen
  \bibfield  {author} {\bibinfo {author} {\bibfnamefont {T.~C.}\ \bibnamefont
  {Hufnagel}}, \bibinfo {author} {\bibfnamefont {C.~A.}\ \bibnamefont {Schuh}},
  \ and\ \bibinfo {author} {\bibfnamefont {M.~L.}\ \bibnamefont {Falk}},\
  }\href@noop {} {\bibfield  {journal} {\bibinfo  {journal} {Acta Mater.}\
  }\textbf {\bibinfo {volume} {109}},\ \bibinfo {pages} {375} (\bibinfo {year}
  {2016})}\BibitemShut {NoStop}%
\bibitem [{\citenamefont {Hays}\ \emph {et~al.}(2000)\citenamefont {Hays},
  \citenamefont {Kim},\ and\ \citenamefont {Johnson}}]{johnson00}%
  \BibitemOpen
  \bibfield  {author} {\bibinfo {author} {\bibfnamefont {C.~C.}\ \bibnamefont
  {Hays}}, \bibinfo {author} {\bibfnamefont {C.~P.}\ \bibnamefont {Kim}}, \
  and\ \bibinfo {author} {\bibfnamefont {W.~L.}\ \bibnamefont {Johnson}},\
  }\href@noop {} {\bibfield  {journal} {\bibinfo  {journal} {Phys. Rev. Lett.}\
  }\textbf {\bibinfo {volume} {84}},\ \bibinfo {pages} {2901} (\bibinfo {year}
  {2000})}\BibitemShut {NoStop}%
\bibitem [{\citenamefont {Szuecs}\ \emph {et~al.}(2001)\citenamefont {Szuecs},
  \citenamefont {Kim},\ and\ \citenamefont {Johnson}}]{johnson01}%
  \BibitemOpen
  \bibfield  {author} {\bibinfo {author} {\bibfnamefont {F.}~\bibnamefont
  {Szuecs}}, \bibinfo {author} {\bibfnamefont {C.~P.}\ \bibnamefont {Kim}}, \
  and\ \bibinfo {author} {\bibfnamefont {W.~L.}\ \bibnamefont {Johnson}},\
  }\href@noop {} {\bibfield  {journal} {\bibinfo  {journal} {Acta Mater.}\
  }\textbf {\bibinfo {volume} {49}},\ \bibinfo {pages} {1507} (\bibinfo {year}
  {2001})}\BibitemShut {NoStop}%
\bibitem [{\citenamefont {Hofmann}\ \emph
  {et~al.}(2008{\natexlab{a}})\citenamefont {Hofmann}, \citenamefont {Suh},
  \citenamefont {Wiest}, \citenamefont {Duan}, \citenamefont {Lind},
  \citenamefont {Demetriou},\ and\ \citenamefont {Johnson}}]{johnson08_1}%
  \BibitemOpen
  \bibfield  {author} {\bibinfo {author} {\bibfnamefont {D.~C.}\ \bibnamefont
  {Hofmann}}, \bibinfo {author} {\bibfnamefont {J.-Y.}\ \bibnamefont {Suh}},
  \bibinfo {author} {\bibfnamefont {A.}~\bibnamefont {Wiest}}, \bibinfo
  {author} {\bibfnamefont {G.}~\bibnamefont {Duan}}, \bibinfo {author}
  {\bibfnamefont {M.}~\bibnamefont {Lind}}, \bibinfo {author} {\bibfnamefont
  {M.~D.}\ \bibnamefont {Demetriou}}, \ and\ \bibinfo {author} {\bibfnamefont
  {W.~L.}\ \bibnamefont {Johnson}},\ }\href@noop {} {\bibfield  {journal}
  {\bibinfo  {journal} {Nature}\ }\textbf {\bibinfo {volume} {451}},\ \bibinfo
  {pages} {1085} (\bibinfo {year} {2008}{\natexlab{a}})}\BibitemShut {NoStop}%
\bibitem [{\citenamefont {Hofmann}\ \emph
  {et~al.}(2008{\natexlab{b}})\citenamefont {Hofmann}, \citenamefont {Suh},
  \citenamefont {Wiest}, \citenamefont {Lind}, \citenamefont {Demetriou},\ and\
  \citenamefont {Johnson}}]{johnson08_2}%
  \BibitemOpen
  \bibfield  {author} {\bibinfo {author} {\bibfnamefont {D.~C.}\ \bibnamefont
  {Hofmann}}, \bibinfo {author} {\bibfnamefont {J.-Y.}\ \bibnamefont {Suh}},
  \bibinfo {author} {\bibfnamefont {A.}~\bibnamefont {Wiest}}, \bibinfo
  {author} {\bibfnamefont {M.}~\bibnamefont {Lind}}, \bibinfo {author}
  {\bibfnamefont {M.~D.}\ \bibnamefont {Demetriou}}, \ and\ \bibinfo {author}
  {\bibfnamefont {W.~L.}\ \bibnamefont {Johnson}},\ }\href@noop {} {\bibfield
  {journal} {\bibinfo  {journal} {PNAS}\ }\textbf {\bibinfo {volume} {105}},\
  \bibinfo {pages} {20136} (\bibinfo {year} {2008}{\natexlab{b}})}\BibitemShut
  {NoStop}%
\bibitem [{\citenamefont {Chen}\ \emph {et~al.}(2012)\citenamefont {Chen},
  \citenamefont {Cheng},\ and\ \citenamefont {Liu}}]{liu12}%
  \BibitemOpen
  \bibfield  {author} {\bibinfo {author} {\bibfnamefont {G.}~\bibnamefont
  {Chen}}, \bibinfo {author} {\bibfnamefont {J.}~\bibnamefont {Cheng}}, \ and\
  \bibinfo {author} {\bibfnamefont {C.}~\bibnamefont {Liu}},\ }\href@noop {}
  {\bibfield  {journal} {\bibinfo  {journal} {Intermetallics}\ }\textbf
  {\bibinfo {volume} {28}},\ \bibinfo {pages} {25} (\bibinfo {year}
  {2012})}\BibitemShut {NoStop}%
\bibitem [{\citenamefont {Liu}\ \emph {et~al.}(2012)\citenamefont {Liu},
  \citenamefont {Li}, \citenamefont {Liu}, \citenamefont {Su}, \citenamefont
  {Wang}, \citenamefont {Li}, \citenamefont {Shi}, \citenamefont {Luo},
  \citenamefont {Wu},\ and\ \citenamefont {Zhang}}]{zhang12}%
  \BibitemOpen
  \bibfield  {author} {\bibinfo {author} {\bibfnamefont {Z.}~\bibnamefont
  {Liu}}, \bibinfo {author} {\bibfnamefont {R.}~\bibnamefont {Li}}, \bibinfo
  {author} {\bibfnamefont {G.}~\bibnamefont {Liu}}, \bibinfo {author}
  {\bibfnamefont {W.}~\bibnamefont {Su}}, \bibinfo {author} {\bibfnamefont
  {H.}~\bibnamefont {Wang}}, \bibinfo {author} {\bibfnamefont {Y.}~\bibnamefont
  {Li}}, \bibinfo {author} {\bibfnamefont {M.}~\bibnamefont {Shi}}, \bibinfo
  {author} {\bibfnamefont {X.}~\bibnamefont {Luo}}, \bibinfo {author}
  {\bibfnamefont {G.}~\bibnamefont {Wu}}, \ and\ \bibinfo {author}
  {\bibfnamefont {T.}~\bibnamefont {Zhang}},\ }\href@noop {} {\bibfield
  {journal} {\bibinfo  {journal} {Acta Mater.}\ }\textbf {\bibinfo {volume}
  {60}},\ \bibinfo {pages} {3128} (\bibinfo {year} {2012})}\BibitemShut
  {NoStop}%
\bibitem [{\citenamefont {Wu}\ \emph {et~al.}(2017)\citenamefont {Wu},
  \citenamefont {Chan}, \citenamefont {Zhu}, \citenamefont {Sun},\ and\
  \citenamefont {Lu}}]{lu17}%
  \BibitemOpen
  \bibfield  {author} {\bibinfo {author} {\bibfnamefont {G.}~\bibnamefont
  {Wu}}, \bibinfo {author} {\bibfnamefont {K.}~\bibnamefont {Chan}}, \bibinfo
  {author} {\bibfnamefont {L.}~\bibnamefont {Zhu}}, \bibinfo {author}
  {\bibfnamefont {L.}~\bibnamefont {Sun}}, \ and\ \bibinfo {author}
  {\bibfnamefont {J.}~\bibnamefont {Lu}},\ }\href@noop {} {\bibfield  {journal}
  {\bibinfo  {journal} {Nature}\ }\textbf {\bibinfo {volume} {545}},\ \bibinfo
  {pages} {80} (\bibinfo {year} {2017})}\BibitemShut {NoStop}%
\bibitem [{\citenamefont {Hamanaka}\ and\ \citenamefont
  {Onuki}(2006)}]{onuki06}%
  \BibitemOpen
  \bibfield  {author} {\bibinfo {author} {\bibfnamefont {T.}~\bibnamefont
  {Hamanaka}}\ and\ \bibinfo {author} {\bibfnamefont {A.}~\bibnamefont
  {Onuki}},\ }\href@noop {} {\bibfield  {journal} {\bibinfo  {journal} {Phys.
  Rev. E}\ }\textbf {\bibinfo {volume} {74}},\ \bibinfo {pages} {011506}
  (\bibinfo {year} {2006})}\BibitemShut {NoStop}%
\bibitem [{\citenamefont {Hamanaka}\ and\ \citenamefont
  {Onuki}(2007)}]{onuki07}%
  \BibitemOpen
  \bibfield  {author} {\bibinfo {author} {\bibfnamefont {T.}~\bibnamefont
  {Hamanaka}}\ and\ \bibinfo {author} {\bibfnamefont {A.}~\bibnamefont
  {Onuki}},\ }\href@noop {} {\bibfield  {journal} {\bibinfo  {journal} {Phys.
  Rev. E}\ }\textbf {\bibinfo {volume} {75}},\ \bibinfo {pages} {041503}
  (\bibinfo {year} {2007})}\BibitemShut {NoStop}%
\bibitem [{\citenamefont {Schun}\ and\ \citenamefont {Lund}(2003)}]{lund03}%
  \BibitemOpen
  \bibfield  {author} {\bibinfo {author} {\bibfnamefont {C.~A.}\ \bibnamefont
  {Schun}}\ and\ \bibinfo {author} {\bibfnamefont {A.~C.}\ \bibnamefont
  {Lund}},\ }\href@noop {} {\bibfield  {journal} {\bibinfo  {journal} {Nat.
  Mater.}\ }\textbf {\bibinfo {volume} {2}},\ \bibinfo {pages} {449} (\bibinfo
  {year} {2003})}\BibitemShut {NoStop}%
\bibitem [{\citenamefont {Ogata}\ \emph {et~al.}(2006)\citenamefont {Ogata},
  \citenamefont {Shimizu}, \citenamefont {Li}, \citenamefont {Wakeda},\ and\
  \citenamefont {Shibutani}}]{ogata06}%
  \BibitemOpen
  \bibfield  {author} {\bibinfo {author} {\bibfnamefont {S.}~\bibnamefont
  {Ogata}}, \bibinfo {author} {\bibfnamefont {F.}~\bibnamefont {Shimizu}},
  \bibinfo {author} {\bibfnamefont {J.}~\bibnamefont {Li}}, \bibinfo {author}
  {\bibfnamefont {M.}~\bibnamefont {Wakeda}}, \ and\ \bibinfo {author}
  {\bibfnamefont {Y.}~\bibnamefont {Shibutani}},\ }\href@noop {} {\bibfield
  {journal} {\bibinfo  {journal} {Intermetallics}\ }\textbf {\bibinfo {volume}
  {14}},\ \bibinfo {pages} {1033} (\bibinfo {year} {2006})}\BibitemShut
  {NoStop}%
\bibitem [{\citenamefont {Zhang}\ \emph {et~al.}(2013)\citenamefont {Zhang},
  \citenamefont {Wang}, \citenamefont {Papanikolaou}, \citenamefont {Liu},
  \citenamefont {Schroers}, \citenamefont {Shattuck},\ and\ \citenamefont
  {O’Hern}}]{ohern13}%
  \BibitemOpen
  \bibfield  {author} {\bibinfo {author} {\bibfnamefont {K.}~\bibnamefont
  {Zhang}}, \bibinfo {author} {\bibfnamefont {M.}~\bibnamefont {Wang}},
  \bibinfo {author} {\bibfnamefont {S.}~\bibnamefont {Papanikolaou}}, \bibinfo
  {author} {\bibfnamefont {Y.}~\bibnamefont {Liu}}, \bibinfo {author}
  {\bibfnamefont {J.}~\bibnamefont {Schroers}}, \bibinfo {author}
  {\bibfnamefont {M.~D.}\ \bibnamefont {Shattuck}}, \ and\ \bibinfo {author}
  {\bibfnamefont {C.~S.}\ \bibnamefont {O’Hern}},\ }\href@noop {} {\bibfield
  {journal} {\bibinfo  {journal} {J. Chem. Phys.}\ }\textbf {\bibinfo {volume}
  {139}},\ \bibinfo {pages} {124503} (\bibinfo {year} {2013})}\BibitemShut
  {NoStop}%
\bibitem [{\citenamefont {Shi}\ \emph {et~al.}(2011)\citenamefont {Shi},
  \citenamefont {Debenedetti}, \citenamefont {Stillinger},\ and\ \citenamefont
  {Ginart}}]{shi11}%
  \BibitemOpen
  \bibfield  {author} {\bibinfo {author} {\bibfnamefont {Z.}~\bibnamefont
  {Shi}}, \bibinfo {author} {\bibfnamefont {P.~G.}\ \bibnamefont
  {Debenedetti}}, \bibinfo {author} {\bibfnamefont {F.~H.}\ \bibnamefont
  {Stillinger}}, \ and\ \bibinfo {author} {\bibfnamefont {P.}~\bibnamefont
  {Ginart}},\ }\href@noop {} {\bibfield  {journal} {\bibinfo  {journal} {J.
  Chem. Phys.}\ }\textbf {\bibinfo {volume} {135}},\ \bibinfo {pages} {084513}
  (\bibinfo {year} {2011})}\BibitemShut {NoStop}%
\bibitem [{\citenamefont {N{\'o}se}(1983)}]{nose83}%
  \BibitemOpen
  \bibfield  {author} {\bibinfo {author} {\bibfnamefont {S.}~\bibnamefont
  {N{\'o}se}},\ }\href@noop {} {\bibfield  {journal} {\bibinfo  {journal} {Mol.
  Phys.}\ }\textbf {\bibinfo {volume} {52}},\ \bibinfo {pages} {255} (\bibinfo
  {year} {1983})}\BibitemShut {NoStop}%
\bibitem [{\citenamefont {Frenkel}\ and\ \citenamefont
  {Smit}(2001)}]{frenkel01}%
  \BibitemOpen
  \bibfield  {author} {\bibinfo {author} {\bibfnamefont {D.}~\bibnamefont
  {Frenkel}}\ and\ \bibinfo {author} {\bibfnamefont {B.}~\bibnamefont {Smit}},\
  }\href@noop {} {\emph {\bibinfo {title} {Understanding Molecular
  Simulation}}},\ \bibinfo {edition} {2nd}\ ed.\ (\bibinfo  {publisher}
  {Academic Press, Oxford},\ \bibinfo {year} {2001})\BibitemShut {NoStop}%
\bibitem [{\citenamefont {Hamanaka}\ \emph {et~al.}(2008)\citenamefont
  {Hamanaka}, \citenamefont {Shiba},\ and\ \citenamefont {Onuki}}]{onuki08}%
  \BibitemOpen
  \bibfield  {author} {\bibinfo {author} {\bibfnamefont {T.}~\bibnamefont
  {Hamanaka}}, \bibinfo {author} {\bibfnamefont {H.}~\bibnamefont {Shiba}}, \
  and\ \bibinfo {author} {\bibfnamefont {A.}~\bibnamefont {Onuki}},\
  }\href@noop {} {\bibfield  {journal} {\bibinfo  {journal} {Phys. Rev. E}\
  }\textbf {\bibinfo {volume} {77}},\ \bibinfo {pages} {042501} (\bibinfo
  {year} {2008})}\BibitemShut {NoStop}%
\bibitem [{\citenamefont {Shiba}\ and\ \citenamefont {Onuki}(2010)}]{onuki10}%
  \BibitemOpen
  \bibfield  {author} {\bibinfo {author} {\bibfnamefont {H.}~\bibnamefont
  {Shiba}}\ and\ \bibinfo {author} {\bibfnamefont {A.}~\bibnamefont {Onuki}},\
  }\href@noop {} {\bibfield  {journal} {\bibinfo  {journal} {Phys. Rev. E}\
  }\textbf {\bibinfo {volume} {81}},\ \bibinfo {pages} {051501} (\bibinfo
  {year} {2010})}\BibitemShut {NoStop}%
\bibitem [{\citenamefont {Allen}\ and\ \citenamefont
  {Tildesley}(1989)}]{allen89}%
  \BibitemOpen
  \bibfield  {author} {\bibinfo {author} {\bibfnamefont {M.~P.}\ \bibnamefont
  {Allen}}\ and\ \bibinfo {author} {\bibfnamefont {D.~J.}\ \bibnamefont
  {Tildesley}},\ }\href@noop {} {\emph {\bibinfo {title} {Computer Simulation
  of Liquids}}},\ \bibinfo {edition} {reprint edition}\ ed.\ (\bibinfo
  {publisher} {Clarendon Press, Oxford},\ \bibinfo {year} {1989})\BibitemShut
  {NoStop}%
\bibitem [{\citenamefont {Halperin}\ and\ \citenamefont
  {Nelson}(1978)}]{nelson78}%
  \BibitemOpen
  \bibfield  {author} {\bibinfo {author} {\bibfnamefont {B.~I.}\ \bibnamefont
  {Halperin}}\ and\ \bibinfo {author} {\bibfnamefont {D.~R.}\ \bibnamefont
  {Nelson}},\ }\href@noop {} {\bibfield  {journal} {\bibinfo  {journal} {Phys.
  Rev. Lett.}\ }\textbf {\bibinfo {volume} {41}},\ \bibinfo {pages} {121}
  (\bibinfo {year} {1978})}\BibitemShut {NoStop}%
\bibitem [{\citenamefont {Nelson}(2002)}]{nelson02}%
  \BibitemOpen
  \bibfield  {author} {\bibinfo {author} {\bibfnamefont {D.~R.}\ \bibnamefont
  {Nelson}},\ }\href@noop {} {\emph {\bibinfo {title} {Defects and Geometry in
  Condensed Matter Physics}}},\ \bibinfo {edition} {1st}\ ed.\ (\bibinfo
  {publisher} {Cambridge University Press, Cambridge},\ \bibinfo {year}
  {2002})\BibitemShut {NoStop}%
\bibitem [{\citenamefont {Yamamoto}\ and\ \citenamefont
  {Onuki}(1997)}]{onuki97}%
  \BibitemOpen
  \bibfield  {author} {\bibinfo {author} {\bibfnamefont {R.}~\bibnamefont
  {Yamamoto}}\ and\ \bibinfo {author} {\bibfnamefont {A.}~\bibnamefont
  {Onuki}},\ }\href@noop {} {\bibfield  {journal} {\bibinfo  {journal} {J.
  Phys. Soc. Jpn.}\ }\textbf {\bibinfo {volume} {66}},\ \bibinfo {pages} {2545}
  (\bibinfo {year} {1997})}\BibitemShut {NoStop}%
\bibitem [{\citenamefont {Yamamoto}\ and\ \citenamefont
  {Onuki}(1998)}]{onuki98}%
  \BibitemOpen
  \bibfield  {author} {\bibinfo {author} {\bibfnamefont {R.}~\bibnamefont
  {Yamamoto}}\ and\ \bibinfo {author} {\bibfnamefont {A.}~\bibnamefont
  {Onuki}},\ }\href@noop {} {\bibfield  {journal} {\bibinfo  {journal} {Phys.
  Rev. E}\ }\textbf {\bibinfo {volume} {58}},\ \bibinfo {pages} {3515}
  (\bibinfo {year} {1998})}\BibitemShut {NoStop}%
\bibitem [{\citenamefont {Thompson}\ \emph {et~al.}(2009)\citenamefont
  {Thompson}, \citenamefont {Plimpton},\ and\ \citenamefont
  {Mattson}}]{thompson09}%
  \BibitemOpen
  \bibfield  {author} {\bibinfo {author} {\bibfnamefont {A.~P.}\ \bibnamefont
  {Thompson}}, \bibinfo {author} {\bibfnamefont {S.~J.}\ \bibnamefont
  {Plimpton}}, \ and\ \bibinfo {author} {\bibfnamefont {W.}~\bibnamefont
  {Mattson}},\ }\href@noop {} {\bibfield  {journal} {\bibinfo  {journal} {J.
  Chem. Phys.}\ }\textbf {\bibinfo {volume} {131}},\ \bibinfo {pages} {154107}
  (\bibinfo {year} {2009})}\BibitemShut {NoStop}%
\bibitem [{\citenamefont {Falk}\ and\ \citenamefont {Langer}(1998)}]{falk98}%
  \BibitemOpen
  \bibfield  {author} {\bibinfo {author} {\bibfnamefont {M.~L.}\ \bibnamefont
  {Falk}}\ and\ \bibinfo {author} {\bibfnamefont {J.~S.}\ \bibnamefont
  {Langer}},\ }\href@noop {} {\bibfield  {journal} {\bibinfo  {journal} {Phys.
  Rev. E}\ }\textbf {\bibinfo {volume} {57}},\ \bibinfo {pages} {7192}
  (\bibinfo {year} {1998})}\BibitemShut {NoStop}%
\bibitem [{\citenamefont {Gao}\ \emph {et~al.}(2006)\citenamefont {Gao},
  \citenamefont {B\l{}awzdziewicz},\ and\ \citenamefont {O'Hern}}]{gao06}%
  \BibitemOpen
  \bibfield  {author} {\bibinfo {author} {\bibfnamefont {G.~J.}\ \bibnamefont
  {Gao}}, \bibinfo {author} {\bibfnamefont {J.}~\bibnamefont
  {B\l{}awzdziewicz}}, \ and\ \bibinfo {author} {\bibfnamefont {C.~S.}\
  \bibnamefont {O'Hern}},\ }\href@noop {} {\bibfield  {journal} {\bibinfo
  {journal} {Phys. Rev. E}\ }\textbf {\bibinfo {volume} {74}},\ \bibinfo
  {pages} {061304} (\bibinfo {year} {2006})}\BibitemShut {NoStop}%
\end{thebibliography}%

\end{document}